\documentclass[10pt,a4paper]{article}
\usepackage{latexsym,amssymb,graphicx,epsfig}
\usepackage{amsthm}
\usepackage{subeqn}
\begin{document}

\thispagestyle{empty}

\title{Classical Orbital Magnetic Moment in a Dissipative Stochastic
System}

\author{N. Kumar\\ Raman Research Institute, Bangalore 560080, India}

\date{}
\maketitle{}

\begin{abstract}
We present an analytical treatment of the dissipative-stochastic
dynamics of a charged classical particle confined bi-harmonically
in a plane with a uniform static magnetic field directed
perpendicular to the plane. The stochastic dynamics gives a steady
state in the long-time limit. We have examined the orbital
magnetic effect of introducing a parametrized deviation $(\eta
-1)$ from the second fluctuation-dissipation (II-FD) relation that
connects the driving noise and the frictional memory kernel in the
standard Langevin dynamics. The main result obtained here is that
the moving charged particle generates a finite orbital magnetic
moment in the steady state, and that the moment shows a crossover
from para- to dia-magnetic sign as the parameter $\eta$ is varied.
It is zero for $\eta=1$ that makes the steady state correspond to
equilibrium, as it should. The magnitude of the orbital magnetic
moment turns out to be a non-monotonic function of the applied
magnetic field, tending to zero in the limit of  an infinitely
large as well as an infinitesimally small magnetic field. These
results are discussed in the context of the classic Bohr-van
Leeuwen theorem on the absence of classical orbital diamagnetism.
Possible realization is also briefly discussed.
\end{abstract}


\vspace{6pt}

The Bohr-van Leeuwen (BvL) theorem \cite{1}--\cite{3} on the
stated absence of orbital diamagnetism for a classical system of
charged particles in equilibrium has been one of the surprises of
physics \cite{4}. The static external magnetic field exerts a
Lorentz force on the moving charged particle, acting at right
angle to its instantaneous velocity (${\bf v}$). While such a
gyroscopic force does no work on the particle, it does induce an
orbital cyclotron motion that subtends an amperean current loop.
The magnetic field associated with this current loop is expected
to be non-zero, and directed oppositely to the externally applied
magnetic field -- the Lenz' law. Hence the expectation of a finite
orbital diamagnetic moment $q/2c(\bf{r}\times \bf{v})$ \cite{5}.
Yet, as is known well, the partition function for a classical
system in equilibrium turns out to be independent of the applied
magnetic field, thus giving a zero orbital magnetic moment. And
this has been the surprise \cite{4}. (The field independence of
the classical partition function derives simply from the fact that
the classical partition function involves integration of the
canonical momentum over an infinite range for any given value of
the conjugate coordinate, and thus the magnetic vector potential
($\bf{A}$) entering the Hamiltonian through minimal coupling
$({\bf p}\rightarrow {\bf p}-\frac{q}{c}{\bf A})$ gets eliminated
through a trivial shift of the momentum variable. This shift is,
however, not allowed for a quantum system because of the canonical
non-commutation involved there. Hence the stated quantum origin of
the orbital magnetic moment in equilibrium -- the Landau orbital
diamagnetism \cite{6}). A remarkably heuristic real-space
explanation for the vanishing of the classical orbital moment was
first suggested by Bohr \cite{1} in terms of a cancellation of the
orbital diamagnetic moment of the completed amperean orbits
(Maxwell cycles) in the bulk interior by the paramagnetic moment
subtended by the incompleted orbits skipping the bounding surface
of the system in the opposite sense. This `{\em edge current}' has
a large arm-length, or leverage and, therefore, can effectively
cancel out the bulk diamagnetic moment. The cancellation has
indeed been demonstrated graphically for a simple planar geometry
\cite{7}. This real space-time picture is consistent with the zero
orbital magnetic moment following from an exact analytical
solution of the classical Langevin dynamics with a white noise
describing the motion of the charged particle confined
harmonically in two dimensions, with a uniform static magnetic
field applied perpendicular to the plane \cite{8}. Here, the
steady state $(t\rightarrow \infty)$  orbital moment indeed
vanishes for the given potential confinement (owing to the spring
constant $k$ of the harmonic potential, providing a soft
boundary). Interestingly, this null result persists in the limit
$k\rightarrow 0$, provided it is taken after taking the limit
$t\rightarrow \infty$. This suggests that in this case the
stochastic particle dynamics has had time enough to sense (i.e.,
be influenced by) the confinement ($k$). On the other hand, the
moment survives to a non-zero value if the order of the two limits
is interchanged. (The effect of these so-called Darwinian limiting
processes is also manifest in the case of the quantum version of
the above Langevin treatment \cite{9}). It is to be noted,
however, that in the case of quantum Langevin equation the orbital
moment tends to zero as the Planck constant is formally reduced to
zero, i.e., in the classical limit, for which the noise term
reduces to a classical white noise which is consistent with a
local Stokes friction constant -- in accord with the second
fluctuation-dissipation (II-FD) theorem \cite{10,11}. This
reasonably suggests to us that it may well be the constraint of
the second fluctuation-dissipation relation that forces the
orbital magnetic moment to vanish in the classical case. This is
further supported by our recent numerical simulation \cite{12}.
Motivated by these observations, we have carried out an exact
analytical calculation of the orbital magnetic moment of a charged
particle confined bi-harmonically in two dimensions with a uniform
static magnetic field applied normal to the xy-plane, but now with
the proviso that the stochastic driving force (noise) is a sum of
two uncorrelated noise terms -- an exponentially correlated term
and a delta-correlated term -- and there is a parametrized $(\eta
- 1)$ deviation from the II-FD relation. Interestingly now, we do
obtain a non-zero orbital magnetic moment in the infinite time
limit -- in the steady state. Moreover, the sign of the orbital
magnetic moment turns out to show a dia- to para-magnetic
crossover as the parameter $\eta$ is tuned through $\eta=1$, where
the moment vanishes. In the following we present an exact
analytical treatment of this dissipative-stochastic system and
discuss the results that follow.

Consider the classical dissipative-stochastic dynamics of a
particle, of charge $-e$ and mass $m$, which is confined
bi-harmonically in the $xy$-plane in the presence of a uniform
static magnetic field $B$ applied perpendicular to the plane. The
governing stochastic (Langevin) equations are
\begin{subequations}
\begin{equation}
m\ddot{x}(t)=-kx(t)-\int_0^t\left(\frac{\Gamma}{t_c}e^{-(t-t')/t_c}+
\Gamma_0\delta(t-t')\right)\dot{x}(t'){\rm
d}t'-\frac{eB}{c}\dot{y}(t)+\xi_x(t)
\end{equation}
\begin{equation}
m\ddot{y}(t)=-ky(t)-\int_0^t\left(\frac{\Gamma}{t_c}e^{-(t-t')/t_c}+
\Gamma_0\delta(t-t')\right)\dot{y}(t'){\rm
d}t'+\frac{eB}{c}\dot{x}(t)+\xi_y(t),
\end{equation}
\end{subequations}
where $\xi_i(t)$ is the noise term with
$\langle\xi_i(t)\rangle=0$, and
\begin{equation}
\langle\xi_i(t)\xi_j(t')\rangle=\delta_{i,j}mk_B
T\left(\frac{\gamma}{t_c}e^{-|t-t'|/t_c}+2\eta\gamma_0\delta(t-t')\right),
\end{equation}
and we are interested in the long-time limit $t\rightarrow
\infty$. Here $i=x,y$, and the angular brackets
$\langle\cdots\rangle$ denote average over realizations of the two
un-correlated noise terms -- one a delta-correlated (white) noise
and the other an exponentially correlated noise with a correlation
time $t_c$. This sum of a white noise and an exponentially
correlated noise, we believe, is the simplest non-Markovian
gaussian process allowed by Doob's theorem \cite{13}. Further,
$(\eta-1)$ parametrizes deviation from the II-FD relation as noted
above.

It is convenient to introduce here the quantities
$\Omega_0=\sqrt{k/m}$ (harmonic oscillator circular frequency),
$\frac{eB}{mc}=\Omega_c$ (cyclotron circular frequency), and
$\frac{\Gamma}{m}=\gamma $ (the frictional relaxation frequency).
We further define the following dimensionless parameters
$\tau=\gamma t$ (dimensionless time), and
$\omega_0=\frac{\Omega_0}{\gamma},
\omega_c=\frac{\Omega_c}{\gamma}$ (dimensionless circular
frequencies).

Following now the 'Landau trick', the two coupled Langevin
equations for the real displacements $x(\tau)$  and $y(\tau)$  as
functions of the dimensionless time $\tau$, can be conveniently
combined into a single Langevin equation for the complex
displacement $z(\tau)\equiv x(\tau)+iy(\tau)$, giving:
\begin{equation}
\ddot{z}(\tau)=-\omega_0^2 z(\tau)-\int_0^\tau
\left(\frac{1}{\tau_c}e^{-(\tau-\tau')/\tau_c}+
\frac{\gamma_0}{\gamma}\delta(\tau-\tau')\right)\dot{z}(\tau'){\rm
d}\tau'+i\omega_c\dot{z}(\tau)+g(\tau)\\
\end{equation}
with $\langle g(\tau)\rangle=0$ and $\langle
g(\tau)g^*(\tau')\rangle=\frac{2k_B T}{(m\gamma^2)}
\left(\frac{1}{\tau_c}e^{-(\tau-\tau')/\tau_c}+
(2\eta)\frac{\gamma_0}{\gamma}\delta(\tau-\tau')\right)$.

Note the complex conjugation $(*)$ that we have introduced in
$\langle g(\tau)g^*(\tau')\rangle$ above, as the same will be
needed in subsequent calculations. Also, we have changed over to
the dimensionless time parameter $(\tau)$ , but have retained the
same symbols for the dynamical variables without the risk of
confusion. Inasmuch as the particle motion along the uniform
magnetic field normal to the xy-plane decouples from that in the
xy-plane, the present model equally well describes a 3-dimensional
system. The orbital magnetic moment can now be re-written as
\begin{equation}
\langle M(\tau)\rangle=\frac{e\gamma}{2c}{\rm Im} \langle
z(\tau)\dot{z}^*(\tau)\rangle.
\end{equation}
It is convenient now to introduce the Laplace
transform
\begin{equation}
\tilde{z}(s)=\int_0^\infty e^{-s\tau}z(\tau)d\tau,\ {\rm and}\
\tilde{\dot{z}}(s)=s\tilde{z}(s),
\end{equation}
with the initial conditions $z(0)=0=\dot{z}(0)$ (without loss of
generality, as we are interested in the steady state (long-time
limit $\tau\rightarrow \infty)$.  We obtain straightforwardly
\begin{equation}
\tilde{z}(s)=\frac{\tilde{g}(s)(1+\tau_c
s)/\tau_c}{\left(s^3+\left(\frac{1}{\tau_c}+
\frac{\gamma_0}{\gamma}-i\omega_c\right)s^2+
\left(\omega_0^2+\frac{1}{\tau_c}+\frac{\gamma_0}{\gamma\tau_c}-
\frac{i\omega_c}{\tau_c}\right)s+\frac{\omega_0^2}{\tau_c}\right)}
\end{equation}
and
\begin{equation}
\tilde{\dot{z}}(s)=\frac{s(1+\tau_c
s)\tilde{g}(s)/\tau_c}{\left(s^3+\left(\frac{1}{\tau_c}+
\frac{\gamma_0}{\gamma}-i\omega_c\right)s^2+
\left(\omega_0^2+\frac{1}{\tau_c}+\frac{\gamma_0}{\gamma\tau_c}-
\frac{i\omega_c}{\tau_c}\right)s+\frac{\omega_0^2}{\tau_c}\right)},
\end{equation}
where we have used $\tilde{\dot{z}}(s)=s\tilde{z}(s)$, with the
initial conditions $z(0)=\dot{z}(0)=0$ as noted above.  In order
to inverse-Laplace transform the expressions above, it is
convenient to introduce the factorized denominator
\begin{eqnarray}
D(s)\hspace{-16pt}&&=\left(s^3+\left(\frac{1}{\tau_c}+
\frac{\gamma_0}{\gamma}-i\omega_c\right)s^2+
\left(\omega_0^2+\frac{1}{\tau_c}+\frac{\gamma_0}{\gamma\tau_c}-
\frac{i\omega_c}{\tau_c}\right)s+\frac{\omega_0^2}{\tau_c}\right)\nonumber\\
&&\equiv(s-s_1)(s-s_2)(s-s_3),
\end{eqnarray}
where $s_i (i=1, 2, 3)$ are the three roots of the cubic
denominator $D(s)$. These roots are readily obtained following the
Cardano procedure. We can then write $\tilde{z}(s)$ and
$\tilde{\dot{z}}(s)$ as partial fractions
\begin{equation}
\tilde{z}(s)=\sum_i\frac{a_i}{s-s_i}
\tilde{g}_z(s),
\end{equation}
and
\begin{equation}
\tilde{\dot{z}}(s)=\sum_i\frac{A_i}{s-s_i}
\tilde{g}_z(s),
\end{equation}
where $\{a_i\}$ are given as solutions of the set of equations
\begin{eqnarray}
&&a_1+a_2+a_3=0\nonumber\\
&&a_1(s_2+s_3)+ a_2(s_3+s_1)+a_3(s_1+s_2)=-1\nonumber\\
&&a_1s_2s_3+a_2s_3s_1+a_3s_1s_2=\frac{1}{\tau_c},
\end{eqnarray}
and similarly, $\{A_i\}$ are given by
\begin{eqnarray}
&&A_1+A_2+A_3=1\nonumber\\
&&A_1(s_2+s_3)+ A_2(s_3+s_1)+A_3(s_1+s_2)=-\frac{1}{\tau_c}\nonumber\\
&&A_1s_2s_3+A_2s_3s_1+A_3s_1s_2=0,
\end{eqnarray}

In terms of the above, we now have the inverse Laplace transforms
as
\begin{subequations}
\begin{equation}
z(\tau)=\sum_ia_i\int_0^\tau
e^{s_i(\tau-\tau')/\tau_c}g(\tau')d\tau'
\end{equation}
{\rm and}
\begin{equation}
\dot{z}(\tau)=\sum_j A_j\int_0^\tau
e^{s_j(\tau-\tau'')/\tau_c}g(\tau'')d\tau''.
\end{equation}
\end{subequations}
With this, the orbital magnetic moment in Eq. (4) turns out to be
\begin{eqnarray}
\langle M(\tau)\rangle\hspace{-16pt} && =
\left(\frac{e}{mc}\right)\left(\frac{k_BT}{\gamma\tau_c^3}\right){\rm
Im}\sum_{i,j=1,2,3}a_iA_j^*\int\!\!\!\!\!\int\limits_0^\tau
\left[e^{s_i(\tau-\tau')/\tau_c}e^{s_j^*(\tau-\tau'')/\tau_c}
e^{-|\tau'-\tau''|/\tau_c}\right.\nonumber\\
&&\left. + 2\eta\frac{\gamma_0\tau_c}{\gamma}
e^{s_i(\tau-\tau')/\tau_c} e^{s_j^*(\tau-\tau'')/\tau_c}
\delta(\tau'-\tau'')\right] d\tau' d\tau'',
\end{eqnarray}
where Im denotes the imaginary part. Straightforward integration
gives the $\tau\rightarrow \infty$ (steady-state) limit for the
orbital magnetic moment as
\begin{eqnarray}
M(\infty)\hspace{-16pt}&& =
\left(\frac{e}{mc}\right)\left(\frac{k_BT}{\gamma\tau_c^3}\right){\rm
Im}\sum_{i,j=1,2,3}a_iA_j^* \left[\left(\frac{1}{\frac{1}{\tau_c}+s_j^*}\right)\right.\nonumber\\
&& \left.
\times\left\{\frac{2}{\tau_c(s^*_j-\frac{1}{\tau_c})(s_i+s_j^*)}+\frac{1}{(s_i-\frac{1}{\tau_c})}\right\}
-2\eta\frac{\gamma_0}{\gamma}\tau_c\left(\frac{1}{s_i+s_j^*}\right)\right].
\end{eqnarray}

After some simplification, the above expression for the orbital
magnetic moment $M(\infty)$ reduces to
\begin{equation}
\mu\equiv \frac{M(\infty)}{(\frac{ek_BT}{mc\gamma\tau_c})}=
-2(\eta-1)\frac{\gamma_0\tau_c}{\gamma}{\rm
Im}\sum_{i,j=1,2,3}a_iA_j^* \left(\frac{1}{s_i+s_j^*}\right)
\end{equation}

In Fig. 1, we have plotted the limiting steady-state value of the
dimensionless orbital magnetic moment $\mu$ against the
dimensionless applied magnetic field
$\beta(=\frac{eB}{mc\gamma})$, where $\eta$ parametrizes the II-FD
violation and is varied over the range 0.5--2.0.  As we are
interested here mainly in the matter of principles, we have made a
simple choice for the dimensionless parameters involved, namely
$\frac{1}{\tau_c}=\omega_0=1$ and $\frac{\gamma_0}{\gamma}=0.5$
(the strength of the white noise relative to the exponentially
correlated noise).

\begin{figure}[h]
\centering\resizebox{10cm}{!}{\includegraphics{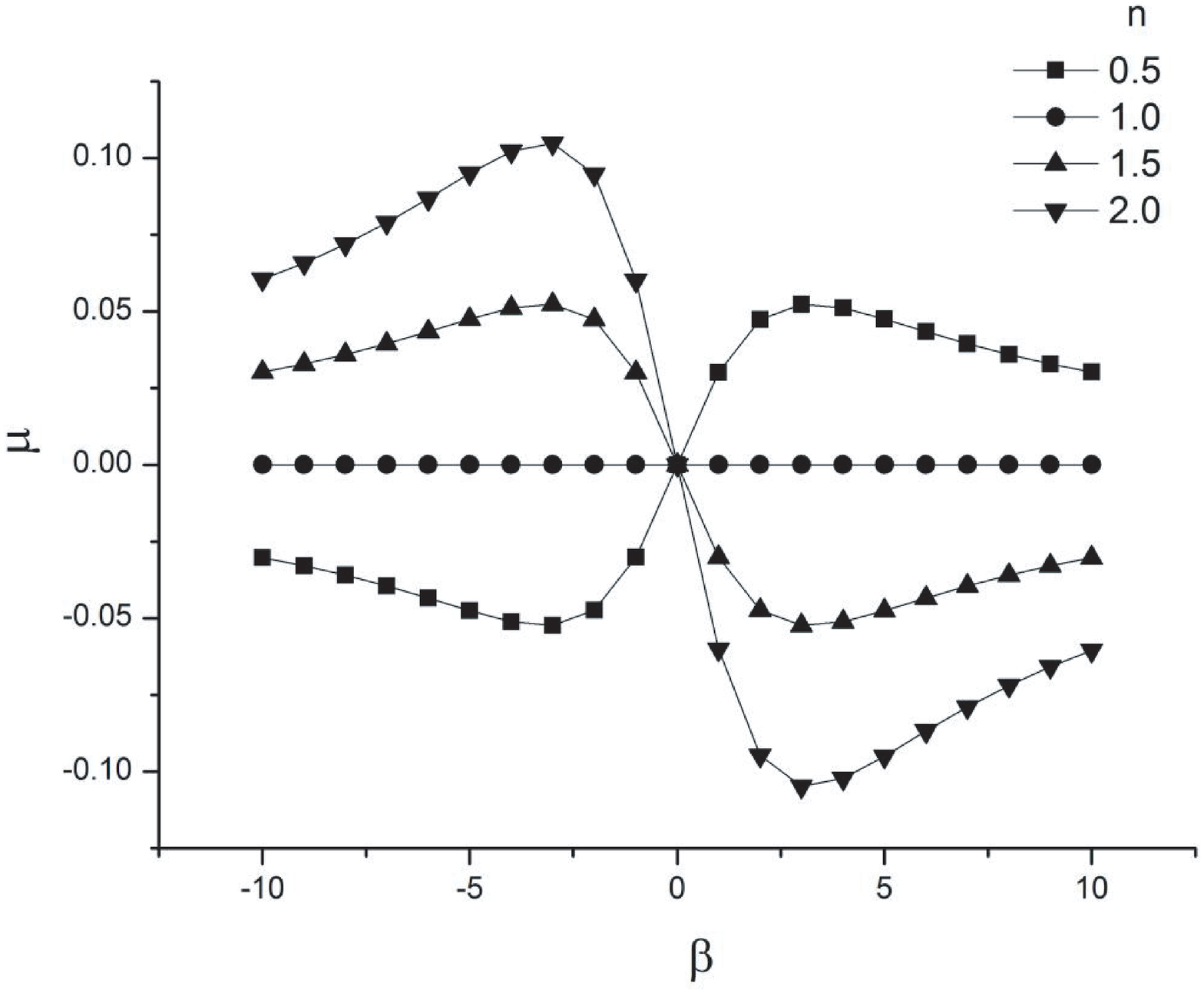}}
\caption{Plot of dimensionless magnetic moment $\mu$ against
dimensionless magnetic field $\beta$ for four different values of
$\eta$ that parametrizes deviation from the II-FD relation. It
clearly shows a dia- to para-magnetic crossover as $\eta$ is
varied through $\eta=1$. Also, the moment tends to zero in the
limit of zero as well as large magnetic field
$\beta$.}\label{fig1}
\end{figure}

As we observe the field-induced orbital magnetic moment is clearly
non-zero in the steady state. There is a crossover from the
paramagnetic to the diamagnetic sign as the parameter $\eta$ is
tuned from $\eta=0.5$ to $\eta=2.0$. This crossover is a surprise.
The magnetic moment is zero for $\eta=1$, which corresponds to the
canonical II-FD consistent (equilibrium) state.  Hence no
violation of the BvL theorem. The orbital magnetic moment is
obviously zero for zero magnetic field; but, not so obviously it
tends to zero for large magnetic field as well. The latter
behaviour can, however, be understood from the following, namely
that the radius/frequency of the cyclotron orbit tends to
zero/infinity as the applied magnetic field is made infinitely
large \cite{4}.  The fact that the orbital magnetic moment can be
paramagnetic in certain range of the II-FD deviation parameter
$\eta$ is significant in that, unlike diamagnetism, it leads to a
positive feedback for a collection of charged particles in such a
classical system -- it can give an enhancement of the orbital
paramagnetism. 

Physical realization of such a classical system in
the laboratory is admittedly somewhat demanding. One needs to
create a dilute (highly non-degenerate) gas of charged particles
(e.g., electrons/holes) at sufficiently high temperatures, and
confined on a mesoscopic scale in the presence of a static
magnetic field. The temperature has to be high enough so as to
wash out the quantum effects, namely the discreteness of the
quantized level spacings owing to the mesoscopic confinement.
Now, the II-FD violating parameter $(\eta-1)$ necessarily
requires a non-equilibrium steady-state condition. This is the real problem for an experimental realization.   One is tempted to think  that such a non-equilibrium steady state may 
be induced through a noisy laser excitation, e.g.,
the Kubo-Anderson non-Markovian noise \cite{14,15}, of charged particles confined in an optical tweezer \cite{16}. There is, however, a problem here involving the energy injected by the laser, its dissipation in the system and the associated rise of temperature. In fact, one necessarily needs to have a two-temperature configuration. Thus, e.g., a possible physical realization of our model can, in principle, have charged particles in contact with a thermal reservoir at one temperature, while another type of neutral particles is in contact with the reservoir at a different temperature.  Then particle collisions will ensure a stationary (steady-state) non-equlibrium state with flow of energy between the reservoirs. This then according to our model calculattion should give a non-zero orbital magnetic moment in an externally applied magnetic field \cite{17}. 

Given that classically a static magnetic field does no
work on a moving charged particle, our model calculation giving 
a non-equilibrium steady-state solution
in the presence of a static magnetic field
could lead to some insightful molecular dynamical (MD)
simulations when appropriately thermostatted \cite{18}. 
Also, it has a significant bearing on the work related to generalized fluctuation-dissipation theorem for steady-state systems \cite{19}.

In conclusion, we have presented an exact analytical treatment of
a classical dissipative-stochastic model system in a uniform
static magnetic field, which is found to give a finite orbital
magnetic moment in the steady state. Interestingly, we find that
there is a crossover from the diamagnetic to the paramagnetic sign
of the magnetic moment as function of a parametrized deviation
from the second fluctuation-dissipation relation. We think that
these results do complement, rather than violate the classic
Bohr-van Leeuwen theorem.

The author would like to thank A.A. Deshpande and K. V. Kumar for
many discussions. The author would also like to thank the referee for constructive criticisms.


\begin{thebibliography}{99}
\bibitem{17}   The author would like to thank the referee for suggesting the possible physical realization of our model.
\end{thebibliography}
\end{document}